\documentclass[%
 reprint,
 amsmath,amssymb,
 aps,
]{revtex4-2}

\usepackage{graphicx}
\usepackage{dcolumn}
\usepackage{bm}
\usepackage[utf8]{inputenc}
\usepackage{multirow} 
\usepackage{xcolor}
\usepackage[colorlinks=true,urlcolor=black,linkcolor=blue,citecolor=blue]{hyperref}

\usepackage{hyperref}
\begin{document}

\preprint{APS/123-QED}

\title{Statistical Estimation of $\mathcal{\pi}$: Varying Choices over Dimensions}

\author{Syon Bhattacharjee$^{1}$ and Subhra Sankar Dhar$^{2}$}
\affiliation{$^{1}$Delhi Public School Kalyanpur, Kanpur, India}
\affiliation{$^{2}$Department of Mathematics and Statistics, Indian Institute of Technology Kanpur, India}

\date{\today}

\begin{abstract}
This article studies statistical estimation of $\pi$ based on the fact that the ratio of the volumes of a $d$-dimensional hypersphere and a $d$-dimensional hypercube is a certain function of $\pi$, and the function depends on the dimension $d$. The estimation of $\pi$ is carried out for various choices of $d$ (strictly speaking, $d\in\{1, 2, \ldots, 20\}$) using the idea of Monte Carlo simulations. Various intriguing facts are observed, and the estimation of $\pi$ using infinite dimensional observations is outlined. Moreover, the R codes associated with relevant numerical studies are provided. 
\end{abstract}

\maketitle
\section{\label{introduction}Introduction}
\subsection{\label{what_is_pi} What is $\mathcal{\pi}$?}
The number $\pi$ is a constant that has intrigued mathematicians for millennia. $\pi$ is defined as the ratio between a circle’s circumference and its diameter. The symbol “$\pi$” was first introduced by William Jones in 1706 and popularized by Leonhard Euler in the 18th century, cementing its place in mathematical notation \cite{arndt2011history}. $\pi$ is not a rational number and cannot be expressed as a fraction of integers. In 1770, Johann Heinrich Lambert first proved its irrationality. Subsequent centuries brought numerous formulas to approximate $\pi$, each differing in accuracy and computational difficulty. The desire for precision and accuracy in approximating $\pi$ was not merely a theoretical pursuit, though; $\pi$ has played a pivotal role in fields ranging from astronomy to architecture, where even minor deviations exponentiate to significant consequences. Thus, accuracy is crucial in calculations where errors can multiply rapidly.  Beyond its role as a constant, $\pi$ embodies a profound metaphor for human inquiry: the further one delves, the richer and more intricate its properties become \cite{galton2009magic}. The Gaussian distribution, which is central to hypothesis testing, approaches the horizontal axis without ever quite reaching it. Yet, the total area under the curve can be accurately calculated by a method of advanced calculus involving double integrals; and it works out to involve the square root of $\pi$. Even today, the quest for the exact value of $\pi$ continues to yield surprises. In January 2024, physicists Arnab Priya Saha and Aninda Sinha of the Indian Institute of Science, discovered an infinite family of $\pi$ formulas rooted in string theory, highlighting that $\pi$ is not a static discovery but a continually unfolding frontier.
\subsection{\label{the_significance_of_pi} The Significance of $\mathcal{\pi}$}
Although most first encounter $\pi$ in the context of circles during geometry class, its applications extend far beyond the classroom. $\pi$ is embedded in natural and physical processes, making it a universal constant with profound implications. In biology, it emerges in the spacing of zebra stripes, leopard spots, and the arrangements of leaves, sunflower seeds and pinecones: patterns also governed by the golden ratio and Fibonacci sequence \cite{santiagopi}. Pi ($\mathcal{\pi}$) appears in nature primarily through circular and cyclical phenomena, such as planetary orbits and the shape of cells, and in patterns related to probability, like river meanders and animal patterns. In mathematics, rose curves (defined by polar equations involving trigonometric functions) depend on $\pi$ to determine the spacing of petals. In physics, $\pi$ is inseparable from oscillatory motion, appearing in formulas for pendulums, harmonic oscillators, and wave propagation. For instance, the period of a simple pendulum is $T=2\pi\sqrt{l/g}$, where $l$ is the length of the pendulum and g is acceleration due to gravity, showing a direct example of how $\pi$ links geometry with temporal dynamics. Even engineering depends on $\pi$: the stability and structural safety of arches, domes, and towers relies on circular properties, while rotating machinery such as turbines and gears requires precise calculations involving angular velocity and torque. In computer science and applied mathematics, $\pi$ plays a vital role in algorithms for integration, random number generation, and numerical simulations. Looking forward, $\pi$ may find novel applications in quantum computing, where mathematical constants play a role in quantum calculations and in artificial intelligence, where its mathematical relationships could influence optimization and pattern recognition \cite{mundase2022pi}. $\pi$ proves itself to be a ubiquitous number, indispensable across scientific, technological, and natural domains. Thus, it is fitting that this universal number is celebrated annually on 14th March - Pi Day.
\subsection{\label{early_historical_approximations_of_pi} Early Historical Approximations of $\mathcal{\pi}$}
 The search for $\pi$’s value dates back nearly four millennia. Ancient Babylonians (c. 1890 B.C.) estimated $\pi$ as 3, with one tablet predicting 3.125. The Egyptians, through the Rhind Mathematical Papyrus (c. 1650 B.C.), approximated it as 3.1605 while solving agricultural field measurement problems. A monumental advance came with Archimedes of Syracuse (c. 287–212 B.C.) who inscribed and circumscribed polygons around a circle, progressively doubling their sides to approximate $\pi$ between 3.1408 and 3.1429 \cite{allen2017approximating}. His principle of exhaustion laid the groundwork for integral calculus and demonstrated that geometry could bind irrational constants with remarkable accuracy. Across Asia, mathematicians made parallel contributions. Liu Hui (3rd century) refined Archimedes’ polygonal method, adopting decimals to yield $\pi = 3.1416$. His successor Zu Chongzhi (5th century) achieved the celebrated approximation of $355/113 \approx 3.1415929$, precise to six decimal places, by working with polygons of up to 24,576 sides. Such achievements illustrate the convergence of independent cultures and traditions toward a shared pursuit of precision.
\subsection{\label{analytical_methods_for_approximating_pi} Analytical Methods for Approximating $\mathcal{\pi}$}
The search for $\pi$ entered a new era with the introduction of infinite series. In the 14th century, the Indian mathematician Madhava of Sangamagrama derived the series: $\pi = 4(1-1/3+1/5-1/7+...)$, which was later rediscovered in Europe as the Gregory---Leibniz series in the 17th century. While conceptually simple, this series converges slowly, requiring many terms for accuracy. In the 17th century, John Wallis introduced his product formula for $\pi/2$, and Newton applied his binomial expansion to arctangent functions, yielding new avenues for computation \cite{bailey1997quest}. These early advances culminated in Machin’s formula: $\pi/4 = 4\arctan (1/5)-\arctan (1/239)$, which provided far faster convergence and became the standard for centuries. In the 19th century, Ramanujan astonished the world with his rapidly converging series, many involving factorials and nested radicals, some producing dozens of correct digits from only a few terms. His work later inspired the modern Chudnovsky algorithm and the Gauss-Legendre method, which are the backbone of current world-record calculations of $\pi$. Remarkably in 2019, Emma Haruka Iwao used Google’s cloud computing to calculate $\pi$ to 31.4 trillion digits, a feat of both mathematical ingenuity and computational power. Each of these methods, whether geometric, analytic, or computational, reflects humanity’s persistent drive to redefine ingenuity.

\subsection{\label{statistical_approaches_for_approximating_pi} Statistical Approaches for Approximating $\mathcal{\pi}$}
Statistical methods, by contrast, offer an easy to comprehend, intuitive and broadly accessible route to approximating $\pi$. By the eighteenth century, new perspectives on $\pi$ had emerged from probability theory. Georges--Louis Leclerc, Comte de Buffon, introduced one of the earliest statistical methods to approximate $\pi$ through his celebrated needle experiment \cite{siniksaran2008throwing}. In this experiment, a needle of length ‘$L$’ is dropped repeatedly onto a floor ruled with parallel lines spaced a distance ‘$d$’ apart. The probability that the needle crosses one of the lines is related to the ratio ‘$2L/\pi d$’. By recording outcomes over many trials, one can estimate $\pi$ from the observed frequency of crossings. This probabilistic approach was revolutionary, as it linked geometric constants to random processes. Centuries later, the development of computational methods provided new opportunities to revisit such ideas and provide alternate proofs. The Monte Carlo method, first formalized in the 20th century, relies on repeated random sampling to approximate quantities that are difficult to compute directly. Estimating $\pi$ is a classic introductory application of this technique. Consider a square of side length ‘$2r$’ centered at the origin with an inscribed circle of radius ‘$r$’. By generating random points $(x,y)$ uniformly across the square and counting the proportion that fall inside the circle, one obtains an estimate of the ratio of the circle’s area to the square’s area. This equality can be rearranged to yield an estimate of $\pi$.
\subsection{\label{proposed_methodology}Proposed Methodology}
Our approach is a systematic investigation using the statistical computing environment of the programming language R. It begins with simulations in two, three, and four dimensions. For each case, we generate between $10^3$ and $10^8$ random sample points uniformly distributed across the unit hypercube, with an inscribed hypersphere (the circle in 2D, sphere in 3D, and corresponding higher-dimensional analogue in 4D). The ratio of points falling inside the hypersphere to the total number of generated points provides a probabilistic estimate of the ratio of the hypersphere’s volume to that of the hypercube, from which $\pi$ can be derived. This method effectively generalizes the familiar circle–inscribed–in–square experiment into higher dimensions. It is extrapolated toward higher--dimensional spaces to test its robustness and scalability. The study seeks to uncover the strengths and limitations of dimensional Monte Carlo methods in approximating $\pi$ \cite{robert2010introducing}. The central objective is thus twofold: to demonstrate the versatility of Monte Carlo simulations for educational and computational purposes, and to understand the relations between dimensionality, sample size, and computational effort.
\subsection{\label{strengths_and_weaknesses_of_the_proposed_approach}
Strengths and Weaknesses of the Proposed Approach}
The proposed methodology offers several notable strengths. First, it demonstrates the fundamental simple principle that accuracy in Monte Carlo estimation improves with larger sample sizes. We can actively control the number of generated points, allowing us to visualize the effect of large numbers in action. Second, the approach is inherently flexible: by modifying the dimension or sample size, one can explore a wide range of experimental conditions without altering the core algorithm. A further strength lies in its extensibility. Additionally, the methodology showcases the versatility of R as a tool for both simulation and visualization. Nevertheless, the method is not without weaknesses. Random number generation itself may introduce variance. This variability means that repeated runs with the same number of points may yield slightly different approximations. Overall, however, the strengths outweigh these limitations. The method remains a powerful and engaging way to approximate $\pi$.
\subsection{\label{organization_of_the_article}Organization of the Article}
The remainder of this paper is organized as follows. Section II elaborates on our proposed methodology. It introduces the computational framework employed, explaining the use of R for random number generation and simulation. It covers experiments in two, three, and four dimensions, as well as extensions towards 20 dimensions. Section III presents the results, analysing accuracy, convergence, and the relations between dimensionality and computational effort. Section IV discusses the concept of infinite dimensional spaces and infinite sample points and concludes the paper by summarizing key findings and outlining potential directions for future research. Section V is the section of References, listing the scholarly sources consulted.

\section{\label{our_methodology}Our Methodology }
\subsection{\label{mathematical_framework}Mathematical Framework}
For an $d$-dimensional hypersphere of radius $r=1$ unit inscribed within a hypercube of side $2r$ unit, the theoretical ratio between their volumes is given by
\begin{equation}
P = \frac{V_{\text{sphere}}}{V_{\text{cube}}}
= \frac{\pi^{d/2}/\Gamma\!\left(\dfrac{d}{2} + 1\right)}{2^{d}},
\end{equation}
where $\Gamma$ is the gamma function ($\Gamma(d) = (d - 1)!$, when $d\in\mathbb{N}$) \cite{li2011concise}. In this formula, for example in 4 dimensions, the volume of a 4–degree hypercube is $2^4$ and that of the 4–degree hypersphere is $\pi^2/\Gamma(3)$. The probability $P$ represents the fraction of randomly generated points within the hypercube that fall inside the hypersphere \cite{weisstein2002hypersphere}. Rearranging this equation allows estimation of $\pi$ as:
\begin{equation}
\pi = \left( P \times 2^{d} \times \Gamma\!\left( \frac{d}{2} + 1 \right) \right)^{\frac{2}{d}}.
\end{equation}
This formula generalizes the classical circle–inside–square method to $d-$dimensional space, linking geometric probability to the value of $\pi$.

\subsection{\label{simulation_design}Simulation Design}
Monte Carlo simulations were performed in R by generating uniformly distributed random coordinates $(x_{1},\, x_{2},\, \ldots,\, x_{n})$ in the interval $[-1, 1]^{d}$. 
\begin{equation}
\{ (x_1, x_2) \;:\; -1 \leq x_1 \leq 1, \; -1 \leq x_2 \leq 1 \}
\end{equation}

The hypersphere and hypercube were both centered at the origin. For each dimension, the distance of the side of the hypercube from the origin was always 1 unit. For each generated point, the squared Euclidean distance from the origin was computed as:

\begin{equation}
x_{1}^{2} + x_{2}^{2} + \cdots + x_{n}^{2} = p^{2}.
\end{equation}
Points satisfying $p^{2} \le 1$ were counted as lying inside the hypersphere \cite{geeksforgeeks_estimating_pi_mc}. The ratio of points inside to the total number of points generated is proportional to the probability of a point lying inside the hypersphere.

Simulations were repeated for sample sizes ranging from $10^3$ to $10^8$ points, depending on the dimension. For almost each configuration, five independent runs were conducted and the average value was determined. Next, the average of the 3 - 5 sample sizes’ values was calculated to finalise the predicted value of $\pi$ for that dimension. The R language was used for all experiments, employing its runif () function for random sampling and vectorized computations to optimize efficiency. Higher–dimensional simulations post 15 dimensions were difficult to carry out on a standard consumer laptop (Lenovo IdeaPad 3 15ITL6 laptop equipped with an 11th Gen Intel® Core™ i3-1115G4 processor @ 3.00 GHz, 8 GB RAM, and a 477 GB SSD). Hence, 2 independent runs were conducted post 15 dimensions for a sample size of $10^8$ points.

\subsubsection{\label{code_design}Code Design}
\begin{figure}
	\centering
    \phantomsection     
    \phantomsection     
	\includegraphics[scale=0.7]{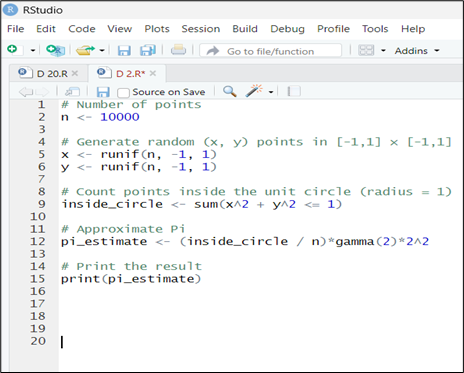}
	\caption{R code to run simulations in 2 dimensions}
    \label{2d_figure1}
\end{figure}

\begin{figure}
	\centering
	\includegraphics[scale=0.7]{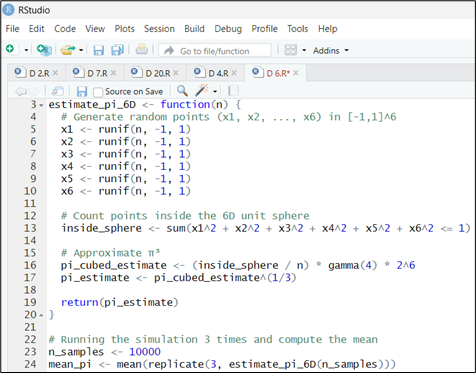}
	\caption{R code to run simulations in 6 dimensions.}
    \label{2d_figure2}
\end{figure}

Figure \ref{2d_figure1} and \ref{2d_figure2} show the code written in R for simulations in 2 dimensions and 6 dimensions.

\subsection{\label{dimensional_extension}Dimensional Extension}
The method was extended from 2 dimensions to 20 dimensions to investigate the effect on the predicted value of $\pi$. In higher dimensions, the volume of the hypersphere shrinks relative to the hypercube, reducing the percentage of points that fall within the sphere. Thus, to maintain a high level of precision, the number of points were increased exponentially. For example, beyond 8 dimensions, $10^3$ points no longer provide reasonable stability and accuracy; beyond 11 dimensions, $10^5$ points were required; beyond 14 dimensions, $10^6$ points were required and beyond 15 dimensions, $10^7$ or more points were necessary.

\section{\label{results_and_analysis}Results and Analysis}
\subsection{\label{convergence_in_lower_dimensions}Convergence in Lower Dimensions}
For 2–D simulations, the estimated values of $\pi$ ranged between 3.056 and 3.208 for 1000 points, converging to an average of 3.1413 at 5000 points and 3.1446 at 10000 points. Similar convergence trends were observed in 3–D, 4–D, 5–D, 6–D and 7–D experiments, where estimates stabilized around 3.14 as the number of sample points reached 10000. As the number of dimensions kept increasing towards 7, the final predicted value of $\pi$ (reported in the 'Final' column) for that dimension showed an increasing trend, as seen from 3.1323 to 3.1404. At 7–D, 1000 points first started to show signs of inaccuracy when one run predicted a value of 2.97, indicating need for increasing the number of points generated. The results for 2–D to 7–D are summarised in Table \ref{tableI}. \\[0.5cm]
\textbf{Note} – The values in the 'Average' column represent the mean of the five values reported in the adjacent five columns (five runs) to the left. The value in the 'Final' column is the mean of the values reported in the 'Average' (preceding) column, and is referred to as the final predicted value of $\pi$' in text.

\begin{table*}[htbp]
\caption{Observations from 2–D to 7–D}
\label{tableI}
\begin{ruledtabular}
\begin{tabular}{cccccccccc}
\textbf{Dim.} & \textbf{Run 1} & \textbf{Run 2} & \textbf{Run 3} & \textbf{Run 4} & \textbf{Run 5} & \textbf{Average} & \textbf{Final} & \textbf{Points} \\
\hline
\multirow{3}{*}{2}
 & 3.116 & 3.056 & 3.060 & 3.116 & 3.208 & 3.111 & \multirow{3}{*}{3.1323} & 1000 \\
 & 3.1456 & 3.1352 & 3.148 & 3.1312 & 3.1464 & 3.1413 &  & 5000 \\
 & 3.148 & 3.1588 & 3.138 & 3.1548 & 3.1232 & 3.1446 &  & 10000 \\[2pt]
\hline
\multirow{3}{*}{3}
 & 3.126 & 3.006 & 3.162 & 3.054 & 3.246 & 3.119 & \multirow{3}{*}{3.1266} & 1000 \\
 & 3.0912 & 3.0504 & 3.0576 & 3.192 & 3.1428 & 3.1068 &  & 5000 \\
 & 3.1326 & 3.1158 & 3.195 & 3.1716 & 3.1554 & 3.1541 &  & 10000 \\[2pt]
\hline
\multirow{3}{*}{4}
 & 3.1496 & 3.1343 & 3.0515 & 3.0620 & 3.1749 & 3.1145 & \multirow{3}{*}{3.1288} & 1000 \\
 & 3.1506 & 3.0797 & 3.0994 & 3.1302 & 3.1230 & 3.1166 &  & 5000 \\
 & 3.1759 & 3.1546 & 3.1066 & 3.1743 & 3.1658 & 3.1554 &  & 10000 \\[2pt]
\hline
\multirow{3}{*}{5}
 & 3.1832 & 3.0835 & 3.0992 & 3.1832 & 3.1531 & 3.1404 & \multirow{3}{*}{3.1390} & 1000 \\
 & 3.1194 & 3.0630 & 3.1332 & 3.1802 & 3.1178 & 3.1227 &  & 5000 \\
 & 3.1599 & 3.137 & 3.119 & 3.135 & 3.219 & 3.154 &  & 10000 \\[2pt]
\hline
\multirow{3}{*}{6}
 & 3.0647 & 3.1823 & 3.2186 & 3.2102 & 3.1362 & 3.1624 & \multirow{3}{*}{3.1504} & 1000 \\
 & 3.1772 & 3.1529 & 3.142 & 3.1353 & 3.1662 & 3.1547 &  & 5000 \\
 & 3.1462 & 3.1208 & 3.1513 & 3.1042 & 3.1483 & 3.1342 &  & 10000 \\[2pt]
\hline
\multirow{3}{*}{7}
 & 2.970 & 3.1190 & 3.2736 & 3.2369 & 3.0763 & 3.1352 & \multirow{3}{*}{3.1404} & 1000 \\
 & 3.1345 & 3.1404 & 3.196 & 3.1748 & 3.0965 & 3.1484 &  & 5000 \\
 & 3.1101 & 3.1842 & 3.1241 & 3.1459 & 3.1244 & 3.1377 &  & 10000 \\
\end{tabular}
\end{ruledtabular}
\end{table*}

\subsection{\label{convergence_in_moderate_dimensions}Convergence in Moderate Dimensions}
For 8–D simulations, the estimated values of $\pi$ ranged between 2.9041 and 3.2674 for $10^3$ points, converging to an average of 3.1406 at $5\times10^3$ points, 3.1339 at $10^4$ points, 3.136 at $5\times10^4$ points and 3.1438 at $10^5$ points. 1000 points showed the least accurate predictions, a trend continuing from 7–D. This emphasised the need for increasing the number of points generated to $5\times10^4$ and $10^5$ points, and treating them as standards for the next few dimensions.  Similar convergence trends were observed in 9–D and 10–D experiments, with final estimates stabilizing towards 3.14. Interestingly at 11–D, $10^4$ points started to predict inaccurate results again (2.9587). Thus, $5\times10^5$ and $10^6$ sample points were added to the standard (fixed) triplet of number of sample points generated (as shown in the right-most column of Table \ref{tableI}, \ref{tableII} and \ref{tableIII}). In 12–D and 13–D, the final predicted value of $\pi$ showed a positive error for the first time (3.1499 and 3.1496). In 14–D, for the case of $10^5$ points generated, lack of precise predictions was observed (2.9966), indicating a need for increasing the number of points generated for 15–D. It is noteworthy that the need for increasing the number of points generated has been exponential: first at 8–D, then at 11–D and now at 14–D. The results obtained for 8–D to 14–D are summarised in Table \ref{tableII}.

\begin{table*}[htbp]
\caption{Observations from 8–D to 14–D}
\label{tableII}
\begin{ruledtabular}
\begin{tabular}{cccccccccc}
\textbf{Dimensions} & \textbf{Run 1} & \textbf{Run 2} & \textbf{Run 3} & \textbf{Run 4} & \textbf{Run 5} & \textbf{Average} & \textbf{Final} & \textbf{Points} \\
\hline
\multirow{5}{*}{8}
 & 2.9041 & 2.925  & 3.1972 & 3.2674 & 3.0785 & 3.0744 & \multirow{5}{*}{3.1257} & 1000 \\
 & 3.1409 & 3.1735 & 3.1968 & 3.0642 & 3.1275 & 3.1406 &  & 5000 \\
 & 3.1942 & 3.0997 & 3.1016 & 3.1762 & 3.0976 & 3.1339 &  & 10000 \\ 
 & 3.1942 & 3.1305 & 3.1556 & 3.1265 & 3.1464 & 3.1506 &  & 50000 \\
 & 3.1487 & 3.1456 & 3.1440 & 3.1635 & 3.1438 & 3.1471 &  & 100000 \\[3pt]
 \hline

\multirow{3}{*}{9}
 & 3.0503 & 3.0841 & 3.1719 & 3.1397 & 3.2240 & 3.1340 & \multirow{3}{*}{3.1343} & 10000 \\
 & 3.1438 & 3.1397 & 3.1046 & 3.1259 & 3.1281 & 3.1284 &  & 50000 \\
 & 3.1438 & 3.1470 & 3.1267 & 3.1562 & 3.1284 & 3.1404 &  & 100000 \\[3pt]
 \hline

\multirow{3}{*}{10}
 & 3.1896 & 3.0739 & 3.0718 & 3.1912 & 3.1907 & 3.1434 & \multirow{3}{*}{3.1365} & 10000 \\
 & 3.0988 & 3.1106 & 3.0997 & 3.1520 & 3.1191 & 3.1160 &  & 50000 \\
 & 3.1292 & 3.1904 & 3.1322 & 3.1628 & 3.1353 & 3.1500 &  & 100000 \\[3pt]
 \hline

\multirow{5}{*}{11}
 & 3.0383 & 3.0604 & 3.0992 & 2.9587 & 3.338 & 3.0989 & \multirow{4}{*}{3.1304} & 10000 \\
 & 3.1520 & 3.1370 & 3.1411 & 3.1851 & 3.0376 & 3.1306 &  & 50000 \\
 & 3.1566 & 3.1673 & 3.1328 & 3.1610 & 3.0803 & 3.1396 &  & 100000 \\
 & 3.1538 & 3.112 & 3.1231 & 3.1508 & 3.1455 & 3.1366  &  & 500000 \\
 & 3.1278 & 3.1629 & 3.1340 & 3.1641 & 3.1328 & 3.1443 &  & 1000000 \\ [3pt]
 \hline

\multirow{3}{*}{12}
 & 3.1475 & 3.1027 & 3.1161 & 3.1889 & 3.2333 & 3.1577 & \multirow{3}{*}{3.1499} & 100000 \\
 & 3.1837 & 3.1378 & 3.1521 & 3.1334 & 3.1423 & 3.1499 &  & 500000 \\
 & 3.1517 & 3.1334 & 3.1203 & 3.1477 & 3.1568 & 3.1420 &  & 1000000 \\[3pt]
 \hline

\multirow{3}{*}{13}
 & 2.7954 & 3.2733 & 3.1583 & 3.2500 & 3.1354 & 3.1225 & \multirow{3}{*}{3.1496} & 10000 \\
 & 3.2025 & 3.1206 & 3.2034 & 3.1754 & 3.1304 & 3.1665 &  & 50000 \\
 & 3.1451 & 3.1818 & 3.1663 & 3.1677 & 3.1377 & 3.1597 &  & 100000 \\[3pt]
 \hline

\multirow{3}{*}{14}
 & 2.9966 & 3.2026 & 3.0966 & 3.1393 & 3.1164 & 3.1103 & \multirow{3}{*}{3.1327} & 100000 \\
 & 3.1543 & 3.2956 & 3.1380 & 3.0440 & 3.0569 & 3.1378 &  & 500000 \\
 & 3.1570 & 3.17350 & 3.1698 & 3.1777 & 3.0713 & 3.1499 &  & 1000000 \\
\end{tabular}
\end{ruledtabular}
\end{table*}

\subsection{\label{behaviour_across_higher_dimensions}Behaviour Across Higher Dimensions}
As dimensionality increased, a greater number of samples were required to achieve comparable accuracy. After running simulations for 15–D, we realised the need for increasing the number of points generated to $5\times10^7$ and $10^8$ as $10^6$ and $5\times10^6$ no longer provided precise predictions. Simulations from 15 to 20 dimensions continued to show an increasing trend (from 3.115 to 3.1975) in final predicted values of $\pi$, as was expected theoretically. From 16–D onwards, $10^8$ samples became the new requirement for stable estimates. Interestingly, the computational capacities of a standard consumer laptop also got tested through these simulations. The processing limitations of the laptop prevented us from running simulations and exploring behaviour beyond 20 dimensions as the system exhibited instability, resulting in frequent freezes or unexpected terminations. The computational cost highlights the trade-off between accuracy and efficiency. Despite the computational burden, the final predicted values consistently approximated 3.14 within $\pm 0.02$ relative error across all dimensions. The results obtained from 15–D to 20–D are summarised in Table \ref{tableIII}.

\begin{table*}[htbp]
\caption{Observations from 15–D to 20–D. \\}
\label{tableIII}
\begin{ruledtabular}
\begin{tabular}{cccccccccc}
\textbf{Dimensions} & \textbf{Run 1} & \textbf{Run 2} & \textbf{Run 3} & \textbf{Run 4} & \textbf{Run 5} & \textbf{Average} & \textbf{Final} & \textbf{Points} \\
\hline

\multirow{3}{*}{15}
 & 3.1103 & 2.8454 & 3.0879 & 3.3284 & 2.9061 & 3.0556 & \multirow{3}{*}{3.115} & 1000000 \\
 & 3.1318 & 3.1300 & 3.1718 & 3.1164 & 3.1176 & 3.1335 &  & 5000000 \\
 & 3.1312 & 3.1770 & 3.1507 & 3.1458 & 3.1782 & 3.1566 &  & 10000000 \\[3pt]
 \hline

\multirow{3}{*}{16}
 & 3.0476 & 3.0849 & 3.0820 & 3.1522 & 3.1198 & 3.0973 & \multirow{3}{*}{3.1307} & 10000000 \\
 & 3.1534 & 3.1156 & 3.1315 & 3.1469 & 3.0918 & 3.1278 &  & 50000000 (2) \\
 & 3.1660 & - & - & - & 3.1670 & 3.1670 &  & 100000000 (3) \\[3pt]
 \hline

\multirow{3}{*}{17}
 & 3.0340 & 3.2670 & 3.1499 & 3.0340 & 3.1823 & 3.1334 & \multirow{3}{*}{3.1373} & 10000000 \\
 & 3.1696 & 3.1362 & 3.1760 & 3.1362 & 3.0844 & 3.1405 &  & 50000000 (2.5) \\
 & 3.1362 & - & - & - & 3.1397 & 3.138 &  & 100000000 (5.5) \\[3pt]
 \hline

\multirow{3}{*}{18}
 & 3.3767 & 3.309 & 3.228 & 3.1264 & 2.9886 & 3.2057 & \multirow{3}{*}{3.1591} & 10000000 \\
 & 3.1904 & 3.2623 & 3.0771 & 3.1264 & 3.1904 & 3.1693 &  & 50000000 (3) \\
 & 3.1147 & - & - & - & 3.0900 & 3.1024 &  & 100000000 (8) \\[3pt]
 \hline

\multirow{3}{*}{19}
 & 3.1807 & 3.1807 & 3.4214 & 3.5706 & 3.5706 & 3.3848 & \multirow{3}{*}{3.1907} & 10000000 \\
 & 3.0141 & 2.8882 & 3.1807 & 3.1068 & 3.1068 & 3.0593 &  & 50000000 (3.5) \\
 & 3.0141 & - & - & - & 3.2423 & 3.1282 &  & 100000000 (10.5) \\[3pt]
 \hline

\multirow{3}{*}{20}
 & 3.298 & 3.298 & 3.298 & 3.298 & 3.298 & 3.298 & \multirow{3}{*}{3.1975} & 25000000 (1) \\
 & 3.298 & 3.298 & 3.298 & 3.077 & 3.37 & 3.21 &  & 50000000 (4) \\
 & 3.2979 & - & - & - & 2.8710 & 3.0845 &  & 100000000 (13) \\
\end{tabular}
\end{ruledtabular}

\textbf{Note} – The result return time is written in brackets next to the number of points generated. The presence of no brackets indicates that the result return time was instantaneous.
\end{table*}
\vspace{0.5cm} 
\subsection{\label{quantitative_trends}Quantitative Trends}

\begin{table}[htbp]
\caption{Table of Relative Errors. \\}
\label{tableIV}
\begin{ruledtabular}
\begin{tabular}{ccc}
\textbf{Dimensions (A)} & \textbf{Final Predicted} & \textbf{Relative Error} \\
 & \textbf{Value of $\pi$} & \\
\hline

\hline
2  & 3.1323 & $-0.0033$ \\
4  & 3.1288 & $-0.0044$ \\
6  & 3.1504 & $+0.0024$ \\
8  & 3.1257 & $-0.0054$ \\
10 & 3.1365 & $-0.0020$ \\
12 & 3.1499 & $+0.0022$ \\
14 & 3.1327 & $-0.0032$ \\
16 & 3.1307 & $-0.0038$ \\
18 & 3.1591 & $+0.0052$ \\
20 & 3.1907 & $+0.0152$ \\
\end{tabular}
\end{ruledtabular}
\textbf{Note} – The average relative error is minimal: 0.00417.
\end{table}

\begin{figure}
	\centering  
	\includegraphics[scale=0.53]{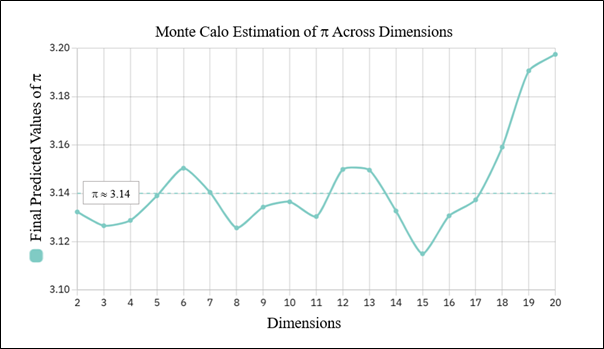}
	\caption{A line graph to show the final predicted values of Pi ($\pi$) for all dimensions (2–D to 20–D). 
    \\ The average of the final predicted values of ($\pi$) for all 20 dimensions is calculated to be 3.1425: a number very close to $\pi$'s theoretically proven value. }
    \phantomsection   
    \label{fig:2d_figure3}
\end{figure}

The observed deviations (as seen in Figure \ref{fig:2d_figure3}) in extremely high dimensions (15–D to 20–D) likely stem from sampling inefficiency due to the sparsity of points inside the hypersphere.

\begin{figure}
	\centering  
	\includegraphics[scale=0.53]{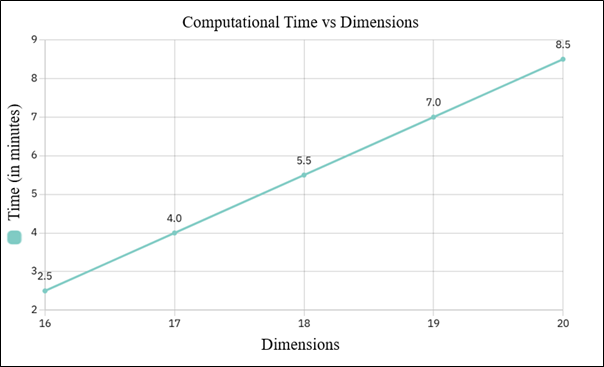}
	\caption{In the following curve, the computational time (in minutes) is plotted against the dimensions. 
    \\ The mean of the times recorded for $5 \times 10^{7}$ and $10^{8}$ sample points was taken as the representative computational time for each dimension. The resulting relationship is linear.}
    \phantomsection   
    \label{fig:2d_figure4}
\end{figure}

\subsection{\label{Computational Complexity}Computational Complexity}

The processing time increased considerably with rise in dimensions (as seen in Figure \ref{fig:2d_figure4}). Until 15–D, all the runs provided instantaneous results. However, from 16–D onwards, the result return time ranged from 2 - 3 minutes initially to even 10 - 13 minutes in higher dimensions.

\section{\label{conclusions}Conclusions}
\subsection{\label{summary_of_findings}Summary of Findings}
In this study, we employed Monte Carlo simulations to approximate the value of Pi ($\pi$) across multiple dimensions using the statistical computing environment of R. By extending the traditional circle–in–a–square experiment to hyperspheres inscribed within hypercubes, the research verified that the ratio of points within the hypersphere to the total number of points converges towards the known value of Pi ($\pi \approx 3.1416$) as the number of random points increases.
 
Dimensional analysis revealed fascinating patterns: lower-dimensional cases (2D--4D) achieved accurate estimates with relatively small sample sizes, whereas higher-dimensional simulations required exponentially more points to maintain comparable precision. These findings are consistent with the law of large numbers and illustrate how convergence depends both on sample size and dimensionality. The results for the final predicted values of Pi ($\pi$) for all even dimensions between 2–D and 20–D are presented in Table \ref{tableIV} along with their relative error. This is also graphically depicted in Fig. \ref{fig:2d_figure3}). 

\subsection{\label{interpretation_and_significance}Interpretation and Significance}

The experiments highlight the robustness and the dual strengths of the Monte Carlo method—it is conceptually simple, as well as adaptable. The code is structured in a manner which eases implementation of the algorithm in R, allowing replication and visualization, while also encouraging further experimentation. The experiments provide evidence allowing us to understand the inherent connection between geometry and probability. The results also offer insight into high-dimensional statistical behaviour, showing how results become more erroneous with increasing dimensionality. 

\subsection{\label{limitations_and_future_improvements}Limitations and Future Improvements}

Despite its strengths, the study acknowledges its limitations. The foremost is computational cost: as both the number of points and the dimensions increase, execution time grows rapidly. Higher-dimensional simulations (higher than 15–D) require millions of iterations and produced greater variability due to the sparsity of points lying inside the hypersphere. Additionally, repeated runs often end up generating the same value or even generate an invalid value of 0.

Future refinements would focus on improving sampling techniques to enhance precision and performance in higher dimensions. Exploring adaptive sampling strategies that allocate more points near hypersphere boundaries might further enhance accuracy.

Overall, the study successfully meets its objectives: it approximates Pi ($\pi$) statistically across 20 dimensions, analyses convergence behaviour, and helps us comprehend the complex web of randomness, precision, and computational efficiency. The findings affirm that Monte Carlo simulations provide a powerful conceptual and computational framework for understanding Pi ($\pi$). The experiments conducted here reinforce the connection between randomness and order: between probability and numerical estimation. The findings demonstrate that even a simple model can possibly unravel deeper mathematical mysteries. 

\subsection{\label{infinite_dimensions_and_infinite_sample Points}Infinite Dimensions and Infinite Sample Points}
While the research demonstrates clear convergence towards $\pi$ in lower and moderate dimensions, it is important to consider the theoretical limits of this methodology. Two natural extensions emerge: the case of infinitely many sample points and the case of infinitely many dimensions. The former explores how the estimator behaves as the number of iterations approaches infinity, while the latter examines the geometric implications of dimensional expansion.

The volume of a $d$-dimensional unit hypersphere is given by

\begin{equation}
V_{d} = \frac{\pi^{d/2}}{\Gamma\!\left( \frac{d}{2} + 1 \right)},
\end{equation}

and the corresponding unit hypercube has a volume of ${2} ^ {d}$. 
\begin{equation}
{As} \ d \to \infty, \quad \frac{V_d}{2^d} \to 0.
\end{equation}

A concise analytical argument supporting this asymptomatic behaviour is outlined below. Stirling’s approximation for the Gamma function is 
\begin{equation}
\Gamma(x+1) \sim \sqrt{2\pi x} \left( \frac{x}{e} \right)^x \cite{mermin1984stirling}.
\end{equation}

Employing Stirling's approximation to the ratio of $V_d$ to $2_d$ as $d \to \infty$ renders

\begin{equation}
\frac{V_d}{2^d} \sim \frac{1}{\sqrt{\pi d}} \left( \frac{\pi e}{2 d} \right)^{d/2}.
\end{equation}

Since ($\frac{\pi e}{2d}) < 1$ for sufficiently large $d$, the ratio decays super-exponentially and tends to 0. Hence, the volume of the unit hypersphere becomes negligible relative to that of the surrounding unit hypercube in higher dimensions.
Intuitively, the hypersphere’s volume initially grows, but eventually shrinks after some dimensions due to the presence of the Gamma function in the denominator. In 2–D, an inscribed circle occupies \(\approx 78.5\%\) of the square's area. In 3-D, an inscribed sphere occupies \(\approx 52.4\%\) of the cube's volume. Thus, as dimensions increase, the hypersphere’s volume shrinks relative to the hypercube - almost all the hypercube is in its “corners”, far from the hypersphere. Additionally, from a tangent viewpoint, distance from the center to a corner of the cube is $\sqrt{d}$; as \(n \to \infty\), the corners are extremely far away relative to the sphere. The hypersphere concentrates near the center, leaving almost all of the cube outside.
Thus, the overall ratio reduces to zero. 
This implies that as dimensionality grows, the probability of a point falling within the hypersphere tends to zero, rendering Monte Carlo estimation unreliable without exponential growth in sample  \cite{tang2024note}.
 
One possible advent of future research involves developing dimension-adaptive Monte Carlo models, in which the number of generated points increases automatically and intelligently with dimensionality to maintain statistical precision. 

\nocite{*}

\bibliography{bibliography}

\end{document}